\documentclass[aps,prl,reprint,superscriptaddress]{revtex4-2}

\usepackage{graphicx}
\usepackage{amsmath}
\usepackage{amssymb}
\usepackage{float}
\usepackage{bigints}
\usepackage[normalem]{ulem}
\usepackage[colorlinks,urlcolor=blue]{hyperref}
\usepackage[capitalise]{cleveref}
\usepackage{xcolor}
\usepackage{mwe}

\begin{document}  
    \title{Local Magnetometry from Measurement-Induced Dissipation}

	\author{Rishith Reddy V	}
    \affiliation{Department of Physics, Indian Institute of Science Education and Research (IISER) Tirupati, Tirupati 517619, India}

    \author{Parveen Kumar}
    \email{parveen.kumar@iitjammu.ac.in}
    \affiliation{Department of Physics, Indian Institute of Technology Jammu, Jammu 181221, India}
	
	\author{Ankur Das}
    \email{ankur@labs.iisertirupati.ac.in}
    \affiliation{Department of Physics, Indian Institute of Science Education and Research (IISER) Tirupati, Tirupati 517619, India}
 
\begin{abstract}
Magnetic phases are commonly identified through macroscopic magnetization, yet many ordered states, including antiferromagnets and altermagnets, possess a vanishing net moment despite distinct local spin structure. We show that such an order can be accessed through the measurement-induced steady state of a single primary qubit locally coupled to a spin lattice. Using a controlled primary–ancillary qubit protocol, we derive analytically that the steady state \emph{encodes} a locally weighted exchange field in a signed observable that is linear in the weak-coupling regime. Numerical simulations demonstrate lattice-scale resolution of antiferromagnetic and altermagnetic textures and robustness against short-correlated noise. Our results establish measurement-induced dissipation as a resource for detecting magnetic order through microscopic structure rather than through global moments.
\end{abstract}	
\maketitle

\emph{Introduction ---}
Magnetism has played a central role in the development of modern condensed matter physics, evolving from early phenomenological observations to a quantum mechanical understanding of collective spin behavior \cite{Coey2020}. This progress has led to a broad classification of magnetic and non-magnetic phases, encompassing ferromagnets and antiferromagnets \cite{Lidiard_1954,PhysRevB.42.4568,Senthil_2004}, as well as more exotic states such as valence bond solids \cite{brezin1990fields,PhysRevB.42.4568,Senthil_2004}, spin liquids \cite{savary2016quantum,Yi_2017,Knolle_2019}, and recently identified altermagnets \cite{PhysRevX.12.040501,Altermagnetism,Bai_2024}. A recurring theme across this diversity is that distinct phases can possess well-defined local spin structure while exhibiting vanishing total magnetization. Identifying and distinguishing such phases, therefore, requires probes that access microscopic magnetic order rather than global/averaged moments.

This requirement poses a fundamental challenge for many established experimental techniques. Probes such as superconducting quantum interference devices (SQUID) magnetometry \cite{Jos_Mart_nez_P_rez_2017} and X-ray magnetic circular dichroism primarily couple to net magnetization and are therefore insensitive to compensated magnetic order \cite{kleiner2004superconducting,van2014x}. Methods with local sensitivity, including spin-polarized scanning tunneling microscopy, offer greater spatial resolution, but are experimentally demanding and often constrained by external magnetic fields \cite{bode2003spin}. These limitations motivate the search for alternative approaches that can detect magnetic order through local spin textures while remaining compatible with a wide range of material platforms.

\begin{figure}
\centering
        \includegraphics[width=1\linewidth]{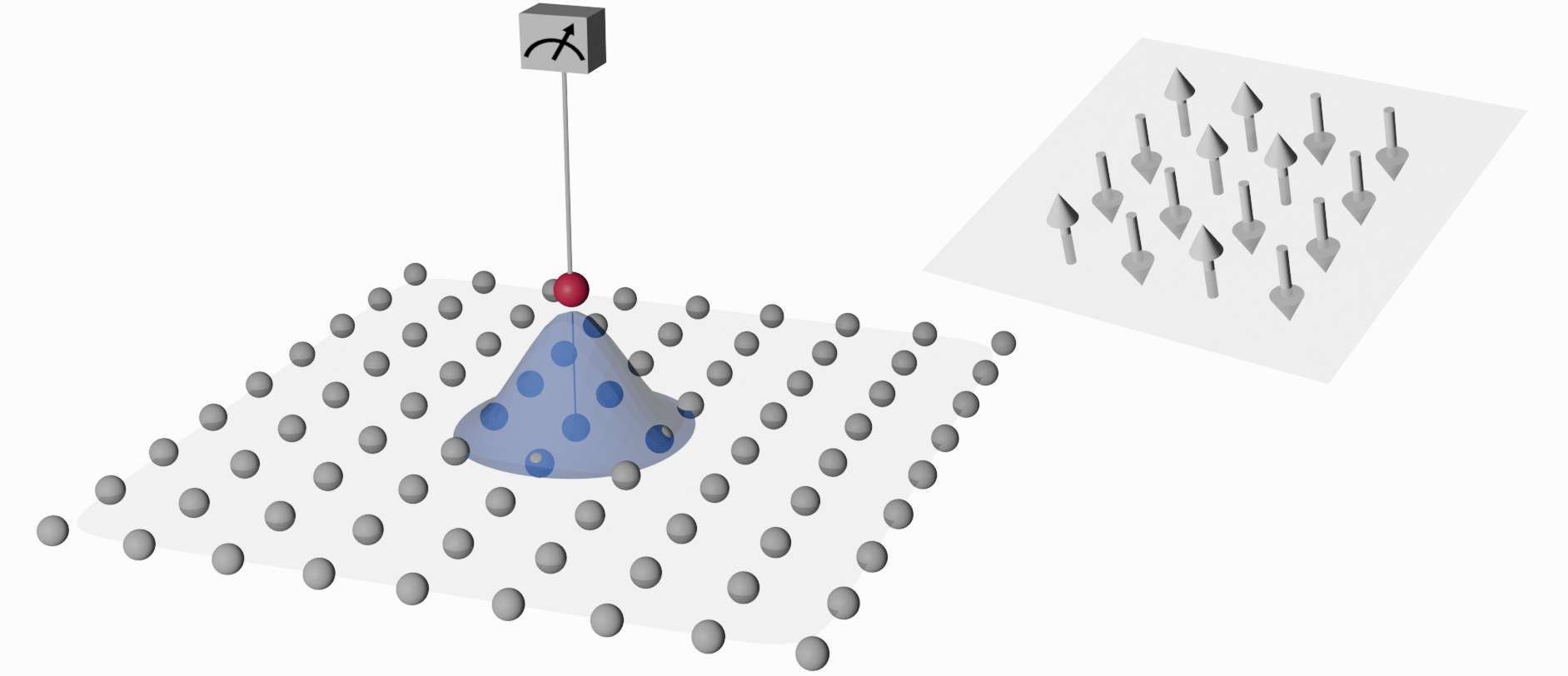}
        \caption{Here a schematic of the setup is presented. The two-level dissipative system (the primary qubit, represented by the red dot) in proximity to a 2D magnetic lattice, coupled via a spin-spin interaction (shown as blue hue) that falls of fast, typical of assumed gaussian coupling. In the inset we show one plausible spin structure of the lattice.}
        \label{fig:Device}
\end{figure}

In this work, we propose a fundamentally different approach based on dissipation induced by measurement of a single two-level quantum sensor (the primary qubit) locally coupled to a spin lattice. The sensor qubit interacts with the surrounding spins through a short-range exchange coupling and is repeatedly coupled to an ancillary qubit that is projectively measured and reinitialized. This measurement protocol generates an effective dissipative dynamics that drives the sensor into a unique steady state~\cite{PhysRevResearch.2.033347,PhysRevResearch.2.042014,PhysRevA.105.L010203,PhysRevResearch.6.023159,PhysRevResearch.6.013244}. The schematic of such a setup is shown in \cref{fig:Device}. We show analytically that this steady state encodes a locally weighted effective magnetic field in a signed observable that is linear in the weak coupling regime, thereby providing direct sensitivity to the local spin configuration rather than to the net magnetization.

Using numerical simulations, we demonstrate that this steady state response enables lattice scale resolution of antiferromagnetic and altermagnetic textures by scanning the sensor across the lattice. Although these phases are indistinguishable to probes that couple only to global magnetization, their distinct local spin arrangements generate different effective fields that are faithfully transduced into the sensor steady state. We further show that this transduction mechanism remains robust in the presence of noise with a short correlation time. Our results establish dissipation induced by measurement as a practical resource for local magnetometry and open the route to detecting magnetic order through microscopic structure rather than bulk moments.

\emph{Model and Setup ---} We consider a two-dimensional spin lattice and a single two-level primary qubit positioned at a location $\mathbf{r}_0$ above the lattice. The primary qubit couples locally to the lattice spins through an exchange-type interaction
\begin{equation}
H_{lp} = \sum_i J_i \, \mathbf{S}_i \cdot \mathbf{S}_p \equiv \mathbf{B}_{\mathrm{eff}} \cdot \mathbf{S}_p ,
\label{eq:Hlp}
\end{equation}
where $\mathbf{S}_i$ denotes the spin at lattice site $i$, $\mathbf{S}_p$ is the spin operator of the primary qubit, and $J_i$ is a distance dependent coupling. Here and in the following, the subscripts $l$, $p$, and $a$ label the lattice, primary qubit, and ancillary qubit degrees of freedom, respectively. Throughout this work, the lattice spins are treated as static classical vectors representing a fixed magnetic configuration, so that the lattice enters the primary qubit dynamics only through the effective field
\begin{equation}
\mathbf{B}_{\mathrm{eff}}(\mathbf{r}_0) = \sum_i J_i(\mathbf{r}_0)\, \mathbf{S}_i .
\label{eq:Beff}
\end{equation}

To model spatial selectivity, we choose a Gaussian coupling profile
\begin{equation}
J_i(\mathbf{r}_0) = \frac{J_0}{d}
\exp\!\left[-\frac{|\mathbf{r}_i - \mathbf{r}_0|^2}{d^2}\right],
\label{eq:Ji}
\end{equation}
where $d$, measured in lattice units, defines an effective sampling radius. For $d \sim 1$, the coupling includes significant nearest neighbor contributions, while smaller values of $d$ correspond to more local probing. For the Ising-type spin configurations considered here, the lattice moments are collinear and the coupling in \cref{eq:Hlp}) reduces to a single-axis interaction of the form $H_{lp} = B_{\mathrm{eff}} \sigma_x / 2$. Although the effective field $\mathbf{B}_{\mathrm{eff}}(\mathbf{r}_0)$ is a vector in general, in this setting, only its projection onto the coupling axis contributes to the qubit dynamics. In the following, we therefore denote by $B_{\mathrm{eff}}$ the relevant scalar component of the effective field.

The sensing protocol is implemented using a measurement-induced dissipative channel constructed from the primary qubit and an ancillary qubit~\cite{PhysRevA.105.L010203}. At the beginning of each cycle, the joint state is taken to be a product
$\rho_{pa} = \rho_p \otimes \rho_a$,
where the ancillary qubit is initialized in a fixed reference state $\rho_a = \tfrac{1}{2}\bigl(\mathbb{I} + \hat{\mathbf{m}}\cdot\boldsymbol{\sigma}\bigr)$. In the following, we choose $\hat{\mathbf{m}} = \hat{z}$, so that $\rho_a = |0\rangle\langle 0|$.

The primary and ancillary qubits undergo a short joint unitary evolution generated by
\begin{equation}
H_{pa} = \frac{J}{2}
\bigl(\boldsymbol{\sigma}_p\cdot\boldsymbol{\sigma}_a
- (\hat{\mathbf{m}}\cdot\boldsymbol{\sigma}_p)
(\hat{\mathbf{m}}\cdot\boldsymbol{\sigma}_a)\bigr),
\label{eq:Hpa}
\end{equation}
after which, the ancillary qubit is projectively measured and reinitialized. Tracing out the ancillary degree of freedom yields an effective map for the primary qubit state
\begin{equation}
\rho_p(t+\mathrm{d}t)
= \sum_i M_i\, \rho_p(t)\, M_i^\dagger ,
\label{eq:Kraus}
\end{equation}
with Kraus operators $M_i = \langle i | U(\mathrm{d}t) | 0 \rangle$, where
$U(\mathrm{d}t) = e^{-i H_{pa} \mathrm{d}t}$.

Retaining terms upto leading order in $\mathrm{d}t$ and taking the continuous time limit with $J^2 \mathrm{d}t \to \alpha$, the measurement protocol generates a purely dissipative contribution to the primary qubit dynamics described by the Lindblad operator $L = |0\rangle\langle 1|$. Combining this with the coherent evolution induced by the lattice coupling, the reduced dynamics of the primary qubit is governed by \cite{openquantumsys}
\begin{equation}
\frac{\mathrm{d}\rho_p}{\mathrm{d}t}
= i[\rho_p, H_{lp}]
- 2\alpha\!\left(
L^\dagger L \rho_p + \rho_p L^\dagger L
- 2 L \rho_p L^\dagger
\right).
\label{eq:ME}
\end{equation}

\cref{eq:ME}) defines the deterministic dynamics of the primary qubit resulting from the competition between coherent precession induced by the lattice and dissipation generated by the measurement protocol. In the following section, we analyze the steady state of this evolution, establish its dependence on the effective field $\mathbf{B}_{\mathrm{eff}}$, and demonstrate how this dependence enables local detection of magnetic order.

\emph{Results ---} We first analyze the steady state of the primary qubit dynamics governed by \cref{eq:ME}. Solving for the steady state, $\mathrm{d}\rho_{p,s}/\mathrm{d}t = 0$, and writing $\rho_{p,s} = \tfrac{1}{2}(\mathbb{I} + \mathbf{s}\cdot\boldsymbol{\sigma})$, we obtain
\begin{equation}
s_x = 0,~~
s_y = -\frac{2\alpha B_{\mathrm{eff}}}{B_{\mathrm{eff}}^2 + 2\alpha^2},~~
s_z = \frac{2\alpha^2}{B_{\mathrm{eff}}^2 + 2\alpha^2}.
\label{eq:steadystate}
\end{equation}
The component $s_y$ provides a signed response to the effective field and is linear in the weak coupling regime $|B_{\mathrm{eff}}| \ll \alpha$. In contrast, $s_z$ depends only on $B_{\mathrm{eff}}^2$ and is therefore insensitive to the field direction. In the following, we use $s_y$ as the primary observable for probing local magnetic order.

We next evaluate $s_y(\mathbf{r}_0)$ by scanning the primary qubit across the lattice. The effective field at each position is given by the locally weighted sum $B_{\mathrm{eff}}(\mathbf{r}_0) = \sum_i J_i(\mathbf{r}_0) S_i$. Although both antiferromagnetic and altermagnetic configurations satisfy $\sum_i \mathbf{S}_i = 0$ and therefore exhibit no net magnetization, their distinct local spin textures generate different spatial patterns of $B_{\mathrm{eff}}(\mathbf{r}_0)$. These differences are directly transduced into the steady state of the primary qubit through \cref{eq:steadystate}.

\begin{figure*}
\centering
\includegraphics[width=\textwidth]{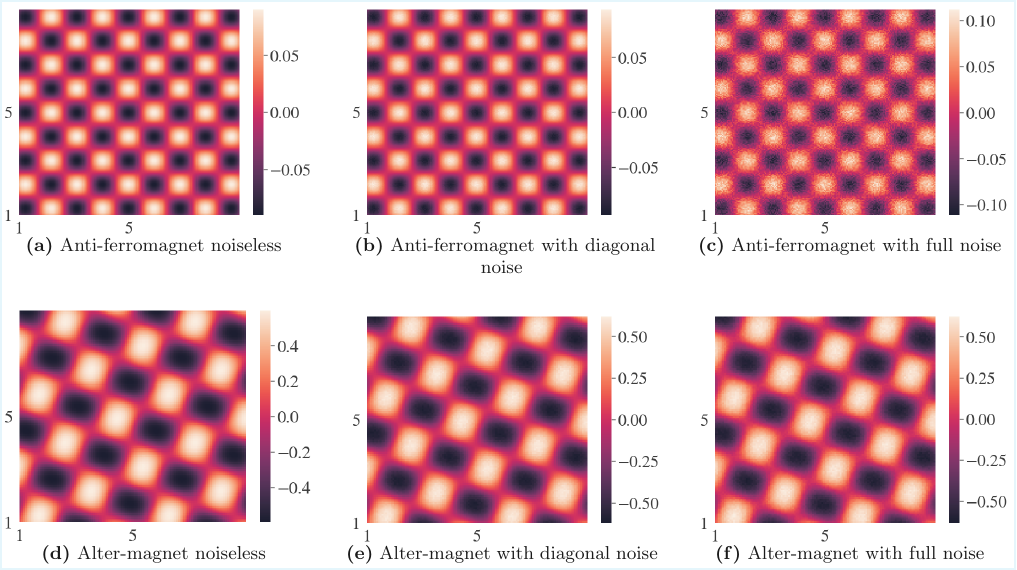} 
    \caption{Spatial maps of the steady-state observable $s_y(\mathbf{r}_0)$ obtained by scanning the primary qubit across the lattice. Panels (a–c) show antiferromagnetic order and (d–f) altermagnetic order \cite{AltermagneticLattice}. Columns correspond to the noiseless case, diagonal noise, and fully correlated noise. Although both phases satisfy $\sum_i \mathbf{S}_i = 0$ and therefore produce no response in probes sensitive only to net magnetization, their distinct local spin textures generate different effective fields $B_{\mathrm{eff}}(\mathbf{r}_0)=\sum_i J_i(\mathbf{r}_0)\mathbf{S}_i$, which are directly reflected in the steady-state response. Noise reduces contrast but preserves the characteristic spatial patterns, demonstrating the robustness of the protocol. Data are shown for $J_0=1$ and $\alpha=1$; noise amplitudes are sampled from $[-0.02,0.02]$.}
    %\PK{Axis ticks and plotlegend ticks are blurred and small. They should be of same size as the text in the paper.}} 
    \label{figure2:main}
\end{figure*}

\cref{figure2:main} shows spatial maps of $s_y(\mathbf{r}_0)$ for both phases. The alternating sign structure characteristic of antiferromagnetic order and the sublattice-dependent patterns of altermagnets are clearly resolved, despite the vanishing global moment in both cases. The contrast arises from the fact that contributions that cancel in the bulk remain finite in the locally weighted sum and are converted into a deterministic steady-state response by the measurement-induced dissipation. 

The spatial resolution is controlled by the sampling radius $d$ in \cref{eq:Ji}. For $d \sim 1$, nearest neighbor contributions dominate and the microscopic structure is sharply resolved. Increasing $d$ leads to stronger spatial averaging, reducing the extrema of $s_y$ without altering the qualitative pattern. This behavior follows directly from \cref{eq:steadystate}, which suppresses the magnitude of the response as $|B_{\mathrm{eff}}|$ decreases while preserving its sign.

To assess robustness, we introduce noise acting on the primary qubit through a stochastic contribution to the Hamiltonian,
\begin{equation}
H_{lp} \rightarrow H_{lp} + \boldsymbol{\xi}(t)\cdot\boldsymbol{\sigma},
\end{equation}
with zero mean and short-time correlations, i.e. $\langle \xi_i(t)\rangle = 0$, and $\langle \xi_i(t') \xi_j(t)\rangle = \gamma_{ij}\,\delta(t'-t)$. Averaging over noise realizations yields an additional Lindblad term in the master equation, whose explicit form depends on the noise correlation matrix~\cite{PhysRevResearch.2.043420}. Solving the resulting equation, we obtain modified steady states of the primary qubit.

For a simpler case when the noise matrix is purely diagonal, i.e. $\gamma_{ij}=0$ for $i\neq j$, the steady state of the primary qubit is given by
\begin{equation}
s_x=0,~ 
s_y=-\frac{2B_{\mathrm{eff}}\alpha}{B_{\mathrm{eff}}^2+2\alpha^2+\beta},~ 
s_z=\frac{2\alpha\left(\alpha+\gamma_{11}+\gamma_{33}\right)}{B_{\mathrm{eff}}^2+2\alpha^2+\beta}.
\end{equation}
where $\beta=2 \alpha \gamma_{33}+\gamma_{22} \left(\gamma_{33}+\alpha\right)+\gamma_{11} \left(\gamma_{22}+\gamma_{33}+3 \alpha\right)+\gamma_{11}^2$. In the case of more general noise, diagonal as well as off-diagonal, the steady state of the physical qubit is given by
\begin{subequations}
\begin{align}
    s_x =& \frac{2\alpha}{\Gamma}\left(\gamma_{12}(\gamma_{23}-B_{\mathrm{eff}}) + \gamma_{13}(\gamma_{11}+\gamma_{33}+\alpha)\right)  \\ 
    s_y =& \frac{2\alpha}{\Gamma}\left(\gamma_{12}\gamma_{13} - (B_{\mathrm{eff}}-\gamma_{23})(\gamma_{22}+\gamma_{33}+\alpha)\right)  \\
    s_z =& \frac{2\alpha}{\Gamma}(\gamma_{11}(\gamma_{22}+\gamma_{33}+\alpha)\notag\\ & \ \ \ \ \ \ \ \ \ \   + (\gamma_{33}+\alpha)(\gamma_{22}+\gamma_{33}+\alpha) - \gamma_{12}^2),
\end{align}
\end{subequations}
% \PK{Formulas in Eq 10, 11, 12, 13 should be written in compact form so that they look nice. Also, we need to define what $\Gamma$ is.}
where $\Gamma = B_{\mathrm{eff}}^2 (\gamma_{22}+\gamma_{33})+3 m^2 \gamma_{22}-m \gamma_{13}^2+m \gamma_{22}^2-m \gamma_{23}^2+\gamma_{33}^2 (\gamma_{22}+2 m)-\gamma_{12}^2 (\gamma_{22}+2 m)+\gamma_{33} ((\gamma_{22}+2 m)^2-\gamma_{13}^2-\gamma_{23}^2)+\gamma_{11}^2 (\gamma_{22}+\gamma_{33}+m)+\gamma_{11} ((\gamma_{22}+\gamma_{33}+m) (\gamma_{22}+\gamma_{33}+3 m)-\gamma_{12}^2-\gamma_{13}^2)-\gamma_{22} \gamma_{23}^2-2 \gamma_{12} \gamma_{13} \gamma_{23}+m (B_{\mathrm{eff}}^2+2 m^2)$.

In the presence of diagonal noise, the amplitude of $s_y$ is reduced, but its spatial structure remains unchanged. Even for full noise, the characteristic patterns associated with antiferromagnetic and altermagnetic order persist, albeit with increased roughness. This behavior reflects the fact that noise suppresses coherence and contrast, while the signed dependence of $s_y$ on $B_{\mathrm{eff}}$, which is linear near the origin, protects the underlying texture from being averaged away.

\emph{Outlook and Future Prospects ---} Our results demonstrate that measurement-induced dissipation can be used as a resource for local magnetometry, enabling detection of magnetic order through microscopic structure rather than through global moments. By converting a locally weighted exchange field into a deterministic steady-state response of a single qubit, the protocol provides direct access to compensated magnetic phases, such as antiferromagnets and alternagnets, which are invisible to conventional probes based on net magnetization.

Beyond the specific examples considered here, the approach is readily extendable. Because the sensing mechanism relies only on local coupling and on repeated primary and ancillary qubit interaction followed by measurement and reset, it can be generalized to probe other observables, including transport properties or correlations, by combining the present protocol with measurement schemes tailored to different operators. Such hybrid strategies offer a route toward multifunctional quantum sensors that access multiple local properties within a unified open-system framework.

Importantly, the ingredients required for the present scheme are compatible with existing experimental architectures. Protocols based on repeated coupling between the primary and ancillary qubit, followed by projective readout and reinitialization, have already been implemented to engineer and control steady states in various platforms, including cold atoms~\cite{bloch2012quantum,diehl2008quantum}, microwave optics~\cite{SHANG2024129263}, Rydberg atoms~\cite{PhysRevA.107.043311}, and superconducting systems~\cite{PhysRevA.88.023849,shankar2013autonomously,PhysRevLett.119.150502}. In particular, a recent experiment with photons has realized measurement-induced steering through primary–ancillary qubit interactions and ancilla reset, demonstrating convergence to predetermined steady states in open quantum dynamics \cite{PhysRevA.110.053717}. These developments suggest that the mechanism underlying our proposal is experimentally accessible, although its application to magnetic sensing remains to be explored. We also mention here that increase in the strength of the measurement, the steady state is achieved faster, which will be important for realistic applications.

More broadly, our work highlights the potential of measurement-induced dissipation as a tool for quantum instrumentation. Exploiting the steady states of open quantum systems rather than transient dynamics opens a pathway to probing ordered phases that are otherwise inaccessible to bulk measurements. We expect that extending this paradigm to interacting sensors~\cite{Gross_2021,asthana2025projectedoptimalsensorsoperator}, multi-qubit architectures \cite{QubitTransportMeasruements}, and time-dependent control will further expand the range of physical phenomena that can be accessed through quantum measurement.

\begin{acknowledgements}
PK acknowledges support from the
IIT Jammu Initiation Grant No. SGT-100106. A.D.\ thanks IISER Tirupati for support and ANRF for start-up grant ANRF/ECRG/2024/001172/PMS.
\end{acknowledgements}

\bibliography{references}

\end{document}